\documentclass[pdftex,twocolumn,epjc3]{svjour3}          

\RequirePackage[T1]{fontenc}

\smartqed  

\RequirePackage{graphicx}
\RequirePackage{mathptmx}      
\RequirePackage{flushend}
\RequirePackage[numbers,sort&compress]{natbib}
\RequirePackage[colorlinks,citecolor=blue,urlcolor=blue,linkcolor=blue]{hyperref}

\journalname{Eur. Phys. J. C}
\begin{document}

\title{Long term measurement of the $^{222}$Rn concentration in the Canfranc Underground Laboratory}

\author
{
J. Amar\'e\thanksref{addr1,addr2}\and I. Bandac\thanksref{addr2}\and A. Blancas\thanksref{addr1}\and S. Borjabad\thanksref{addr2}
\and S. Buis\'an\thanksref{addr3}\and S. Cebri\'an\thanksref{addr1,addr2,addr5}\and D. Cintas\thanksref{addr1,addr2}\and I. Coarasa\thanksref{addr1,addr2}\and E. Garc\'ia\thanksref{addr1,addr2,addr5}\and M. Mart\'inez\thanksref{addr1,addr2,addr4}
\and R. N\'u\~nez--Lagos\thanksref{addr5}\and M.A. Oliv\'an\thanksref{addr1}\and Y. Ortigoza\thanksref{addr5,addr6}
\and A. Ortiz de Sol\'orzano\thanksref{addr1,addr2,addr5}\and C. P\'erez\thanksref{addr1,addr2,addr5}\and J. Puimed\'on\thanksref{e1,addr1,addr2,addr5}\and S. Rodr\'iguez\thanksref{addr5}\and A. Salinas\thanksref{addr1,addr2}\and M.L. Sarsa\thanksref{addr1,addr2}\and P. Villar\thanksref{addr1}
}

\thankstext{e1}{e-mail: puimedon@unizar.es}

\institute{Centro de Astropartículas y Física de Altas Energías (CAPA), Universidad de Zaragoza, Pedro Cerbuna 12, 50009 Zaragoza, Spain\label{addr1}
\and Laboratorio Subterráneo de Canfranc, Paseo de los Ayerbe s/n, 22880 Canfranc Estaci\'on, Huesca, Spain\label{addr2}
\and AEMET, Delegaci\'on Territorial de AEMET en Arag\'on, Paseo del Canal 17, 50007 Zaragoza, Spain\label{addr3}
\and Fundaci\'on ARAID, Avenida de Ranillas 1D, 50018 Zaragoza, Spain\label{addr4}
\and Laboratorio de Bajas Actividades (LABAC), Universidad de Zaragoza, Pedro Cerbuna 12, 50009 Zaragoza, Spain\label{addr5}
\and EUPLA, Calle Mayor 5, 50100 La Almunia de Do\~na Godina, Zaragoza, Spain\label{addr6}
}

\date{Received: date / Accepted: date}

\titlerunning{Long term measurement of the $^{222}$Rn concentration in the Canfranc Underground Laboratory}
\maketitle

\begin{abstract}
We report the results of six years (2013-2018) of measurements of $^{222}$Rn air concentration, relative humidity, atmospheric pressure and temperature in the halls A, B and C of the Canfranc Underground Laboratory (LSC). We have calculated all the Pearson correlation coefficients among these parameters and we have found a positive correlation between the $^{222}$Rn concentration and the relative humidity. Both correlated variables show a seasonal periodicity. The joint analysis of laboratory data and four years (2015-2018) of the meteorological variables outside the laboratory shows the correlation between the $^{222}$Rn concentration and the outside temperature.
The collected information stresses the relevance of designing good Rn-mitigation strategies in current and future experiments at LSC; in particular, we have checked for two years (2017-2018) the good performance of the mitigation procedure of the ANAIS--112 experiment. Finally, in another measurement (2019-2021) for two years of live time, we report an upper limit to the residual $^{222}$Rn content of the radon-free air provided by the radon abatement system installed in the laboratory.
\end{abstract}

\section{Introduction}

Monitoring the level of $^{222}$Rn in laboratories of low activity measurements and, in particular, in underground laboratories is essential to ensure the safety of the workers inside the laboratory. The main remediation action is a correct ventilation \cite{Arpesella1997healthp,BASSIGNANI1997rm,LESKO2015taup2013,liu2018arxiv} to evacuate the radon emanation from the rock and to keep a concentration similar to that of the open air, that usually is <~100~Bq~m$^{-3}$ \cite{Euratlas}.
The posterior actions to provide a very low radioactive background to the experiments in the laboratory depend on their requirements. The experiments can be purged with nitrogen (from compressed nitrogen or evaporated from a liquid nitrogen dewar) or with radon-free air provided that the laboratory has an adequate system to satisfy this demand, see for instance \cite{perezperez2021radon}. Monitoring the $^{222}$Rn level is also very important to avoid the plate-out of the $^{222}$Rn daughter, specially $^{210}$Pb and its descendants,
during the assembly of the components of an experiment \cite{STEIN2018nima}.
The seasonal modulation of the $^{222}$Rn concentration affects in particular to experiments aiming at the study of the annual modulation expected in the interaction rate of the galactic halo dark matter particles \cite{modulacion_Drukier1986}. Because of that, these experiments \cite{lee2011radon,prd-minos-cogent} are specifically isolated from the laboratory air and conveniently purged with radon free gas.

We address three items in this article: monitoring the $^{222}$Rn concentration inside the LSC and its correlation with the internal and external temperature, pressure and relative humidity; the measurement of the residual $^{222}$Rn concentration of the evaporated nitrogen that was used to purge the ANAIS--112 experiment during 2017-2018 \cite{Amare:2018sxx,Amare2021prd} and, finally, a similar measurement of the air provided by the radon abatement system \cite{perezperez2021radon} that supplies radon-free air to the experiments in the LSC \cite{DarT_Garcia_2020,cross_Bandac2020,next_Adams2021,Castel2019,BabyIaxo2021,clyc2018,HENSA-ANAIS_TAUP2021}.
Our data add very useful information to the published results of other background sources at LSC: the rock radioactivity \cite{rock-LSC-taup2005}, the neutron flux \cite{CARMONA2004523,neutrons_LSC-2013} and the cosmic-ray muon flux \cite{muons_LSC-2019}.

\section{Description of the Canfranc Underground Laboratory}
The Canfranc Underground Laboratory, LSC\footnote{Detailed information at www.lsc-canfranc.es.} (Laboratorio Subterráneo de Canfranc) is a Spanish scientific installation located in Arag\'on, in the village Canfranc-Estaci\'on (Huesca) at 1120~m above sea level (geographical coordinates: 42º 46’ 30’’ N; 0º 31’ 42’’ W). Its laboratories and rooms have been excavated in the rock at 800 meters depth, below the Tobazo Mountain, at the Spanish Pyrenees between the Somport road tunnel, which links Spain with France, and the parallel railway tunnel today in disuse.

\begin{figure}
\begin{minipage}{\columnwidth}
\centering
\includegraphics[width=1.00\textwidth]{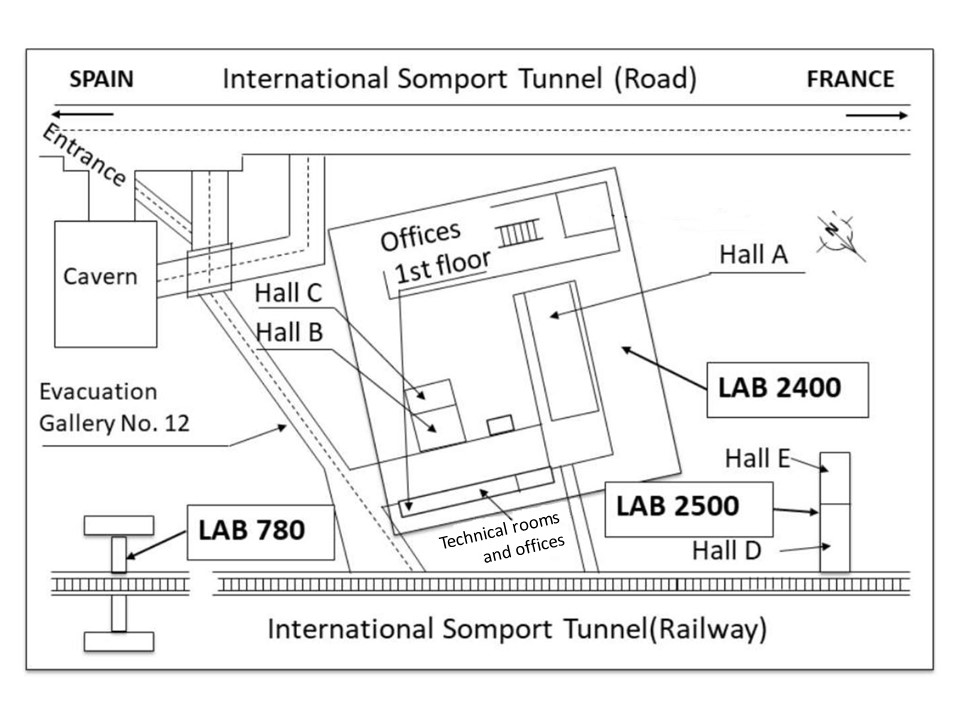}
\end{minipage} 
\caption{Scheme (not to scale) of the Canfranc Underground Laboratory facilities}
\label{fig:LSC_plano}
\end{figure}

The underground infrastructure has a total surface of\linebreak[4] about 1600~m$^{2}$ divided in LAB780 (two small experimental rooms at 260 meters depth), LAB2400 and LAB2500 (both at 800 meters depth), named according to the distance in meters from each location to the Spanish entrance of the railway tunnel (780, 2400 and 2500~m, respectively). This work has focused on the main experimental area (LAB2400) that consists of a large room, hall~A, 40$\times$15$\times$12~(height)~m$^{3}$ and a smaller volume of 15$\times$10$\times$8~(height)~m$^{3}$ divided into two contiguous rooms, halls~B and~C (Figure~\ref{fig:LSC_plano}).

The ventilation provides around 11000 m$^{3}~$h$^{-1}$ of fresh air. The flow of fresh air is taken from the mountain at 1400~m altitude and then reaches an intermediate ventilation room at 1120~m through a ventilation shaft and a ventilation gallery excavated in the rock. From there it is sent to the air conditioning room by a forced ventilation duct, where it joins 14000 m$^{3}~$h$^{-1}$ of the returned air and its final temperature is regulated by the air conditioning system. The fresh air flow guarantees the total renewal of the laboratory air once per hour \cite{Radiopro2014}.

\section{$^{222}$Rn concentration in six years of data}
\label{sec:Rn six years data}

The here reported measurements of the $^{222}$Rn concentration in air (Rn$_{\mathrm{i}}$), the temperature (T$_{\mathrm{i}}$), the atmospheric pressure (P$_{\mathrm{i}}$) and the relative humidity (RH$_{\mathrm{i}}$) have been obtained using three Genitron (now BERTIN) ALPHAGUARD P30\texttrademark\ monitors from 2013 to 2018.
The measurements were taken continuously by a dedicated software that recorded every 10 minutes the values of the four variables and the statistical uncertainty of the $^{222}$Rn concentration.

The calibration uncertainties quoted in the data sheet are 3\% for the $^{222}$Rn concentration activity, 1.5~°C for the temperature, 3~hPa for the atmospheric pressure and 3\% for the relative humidity. In order to check for small variations, each monitor measured in the same hall for the six years. The calibrations of the monitors were tested
by an interlaboratory comparison in 2015 \cite{laruc2015}. The mean departure of our monitors was a 9\% with respect to the $^{222}$Rn concentrations measured by the reference monitor of the comparison.

The $^{222}$Rn annual average activity concentrations in the three halls are listed in Table~\ref{tab:222rn_annos_mean}; their statistical uncertainties are about 0.07\% because the response of the monitor is (5~counts per minute)/(100~Bq~m$^{-3}$). The radon annual level is independent of the hall (within the 3\% instrument calibration uncertainty), it complies with the European directive on radon in workplaces \cite{directive_euratom}, with the ICRP recommendations \cite{icrp103} and it is well below the 600 Bq~m$^{-3}$ allowed by Spanish legislation \cite{csn11-04}. 

\begin{table}
\caption{Annual average activity concentration of $^{222}$Rn (Bq~m$^{-3}$) from 2013 to 2018. Their statistical uncertainties are approximately 0.07\% whereas the instrument calibration error is 3\%. The last column is the annual average for the 6 years of data}
\label{tab:222rn_annos_mean}
\centering
\begin{tabular}{lccccccc}
\hline
Room	&2013&2014&2015&2016&2017&2018&2013-8\\
\hline
Hall A &87.5&82.4&82.8&80.4&80.8&82.9&82.8 \\
Hall B &89.1&83.8&84.0&82.8&84.8&87.0&85.2 \\
Hall C &89.2&82.7&86.1&82.9&83.3&84.4&84.8 \\
\hline
\end{tabular}
\end{table}

\subsection{Correlations between the variables inside the laboratory}
\label{sub:correla_inside}

The relations between meteorological parameters and levels of $^{222}$Rn have been studied before at shallow depth, see for instance \cite{Pujol_Garcia-Tobar_2014}. In this section, we give the results for all possible correlations between every pair of the monthly averages at LSC ($^{222}$Rn concentration, temperature, atmospheric pressure and relative humidity).
Table~\ref{tab:pearsonLSC} shows the values of the Pearson correlation coefficient, $r$, for the monthly averages of all the pairs of parameters measured at halls A, B and C.

\begin{table*}[htbp]
\caption{Values of Pearson correlation coefficient (2013-18) for all pairs of the measured monthly averaged parameters inside the LSC}
\label{tab:pearsonLSC}
\centering
\begin{tabular*}{\textwidth}{@{\extracolsep{\fill}}llrrrrrrrrrrrr@{}}
\hline
& & \multicolumn{3}{c}{Radon} & \multicolumn{3}{c}{Temperature} & \multicolumn{3}{c}{Pressure}& \multicolumn{3}{c}{Relative humidity}\\
\cline{3-5} \cline{6-8} \cline{9-11} \cline{12-14}
   &        & Hall A & Hall B & Hall C & Hall A & Hall B & Hall C & Hall A & Hall B & Hall C & Hall A & Hall B & Hall C\\
\hline
   & Hall A & 1.00 \\
Rn$_{\mathrm{i}}$ & Hall B & 0.98 & 1.00 \\
   & Hall C & 0.96 & 0.94 & 1.00 \\
\cline{1-5}
   & Hall A & $-$0.13 & $-$0.07 & $-$0.16 & 1.00 \\
T$_{\mathrm{i}}$  & Hall B & 0.09 & 0.18 & 0.05 & 0.53 & 1.00 \\
   & Hall C & $-$0.02 & 0.09 & $-$0.03 & 0.32 & 0.86 & 1.00 \\
\cline{1-8}
   & Hall A & 0.11 & 0.13 & 0.11 & 0.14 & 0.12 & 0.06 & 1.00 \\
P$_{\mathrm{i}}$  & Hall B & 0.14 & 0.13 & 0.13 & 0.07 & 0.01 & $-$0.10 & 0.97 & 1.00 \\
   & Hall C & 0.10 & 0.12 & 0.15 & 0.07 & 0.10 & 0.09 & 0.97 & 0.94 & 1.00 \\
\cline{1-11}
   & Hall A & 0.73 & 0.69 & 0.69 & $-$0.14 & 0.22 & 0.05 & 0.34 & 0.37 & 0.36 & 1.00 \\
RH$_{\mathrm{i}}$ & Hall B & 0.72 & 0.68 & 0.67 & 0.00 & 0.12 & $-$0.12 & 0.35 & 0.41 & 0.33 & 0.96 & 1.00 \\
   & Hall C & 0.70 & 0.67 & 0.70 & 0.01 & 0.17 & $-$0.07 & 0.36 & 0.41 & 0.38 & 0.96 & 0.98 & 1.00 \\
\hline	
\end{tabular*}
\end{table*}

The statistical significance of the $r$--value of $n$ pairs of two variables can be easily calculated because the variable
\begin{equation}
	t=r\sqrt{\frac{n-2}{1-r^{2}}} \label{student}
\end{equation}

\noindent follows the Student's $t$ distribution with $n-2$ degrees of freedom for two uncorrelated normal variables \cite{Kendall_1}. Even in the case of non--normal variables or small size $n$, the Student's $t$--distribution is a very good approximation for the $t$ variable in Eq.~(\ref{student}) \cite{Kendall_2}. Therefore, the decision threshold, $\left|r_{\alpha}\right|$, to reject the null hypothesis (no correlation) is obtained from the corresponding two-tailed $\left|t_{\alpha}\right|$.

We take a 99.92\% CL for the decision threshold because if the variables of Table \ref{tab:pearsonLSC} are uncorrelated, the probability to get some $r$--value $\ni$ $\left|r\right|>\left|r_{\alpha}\right|$ with 66 evaluated $r$--values is $1-(0.9992)^{66}=0.05$, similar to a 95\% CL if only one $r$--value were evaluated. The monthly average 6-year data gives $n=72$, i.e., 70 degrees of freedom for the Student's $t$ and $\left|r_{\alpha=0.0008}\right|=0.39$.

\begin{figure*}
\begin{minipage}{\columnwidth}
\centering
\includegraphics[width=2.00\textwidth]{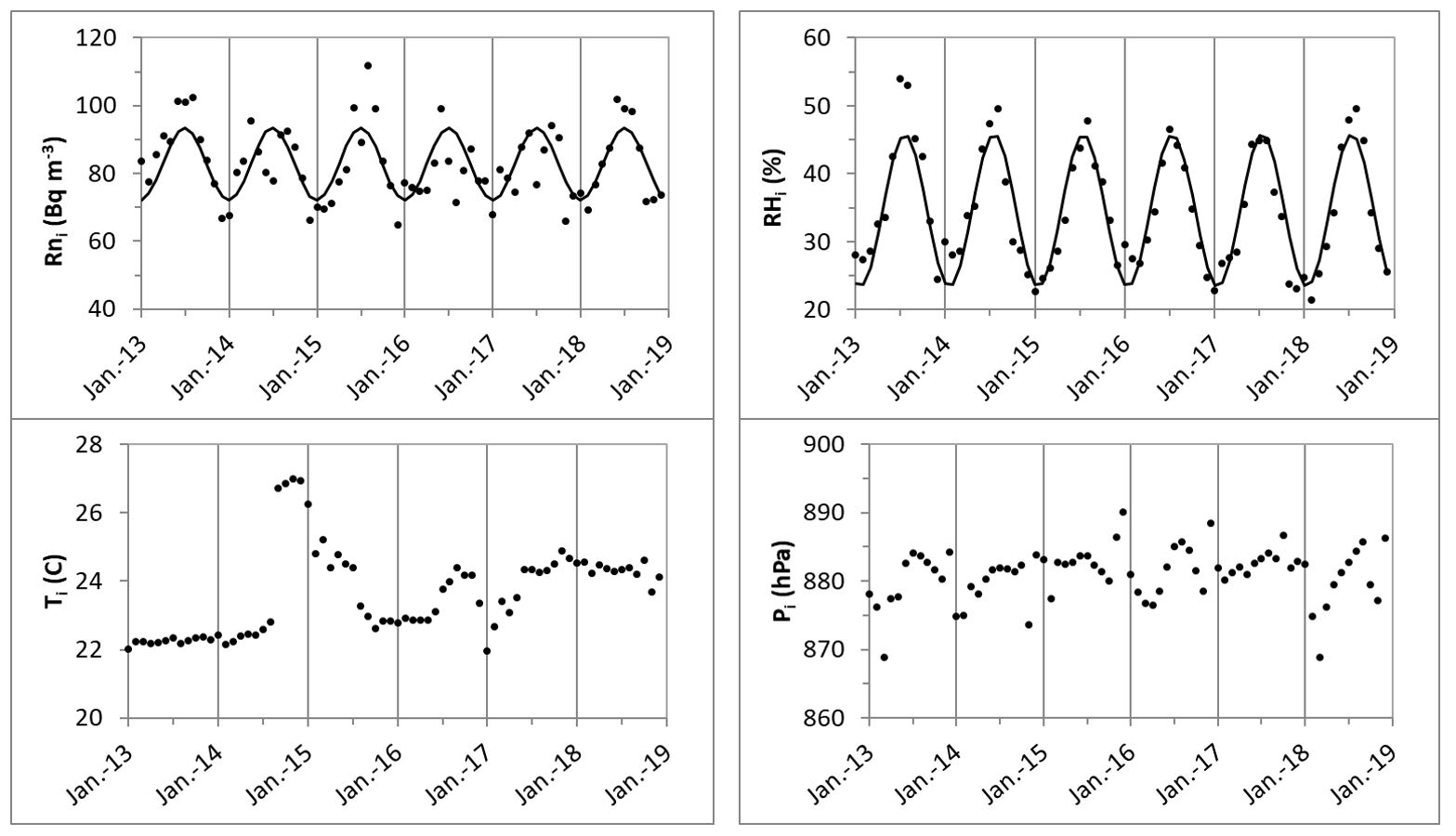}
\end{minipage}
\caption{Monthly average of the hall A data: $^{222}$Rn concentration (upper left), relative humidity (upper right), temperature (bottom left) and atmospheric pressure (bottom right) from January 2013 to December 2018. The fits of the $^{222}$Rn concentration and the relative humidity to Eq. \ref{coseno} are also shown. Corresponding plots for halls B and C are similar, except the temperature (see Figure~\ref{fig:T-halls_ByC})}
\label{fig:datos-halla}
\end{figure*}

\begin{figure*}
\begin{minipage}{\columnwidth}
\centering
\includegraphics[width=2.00\textwidth]{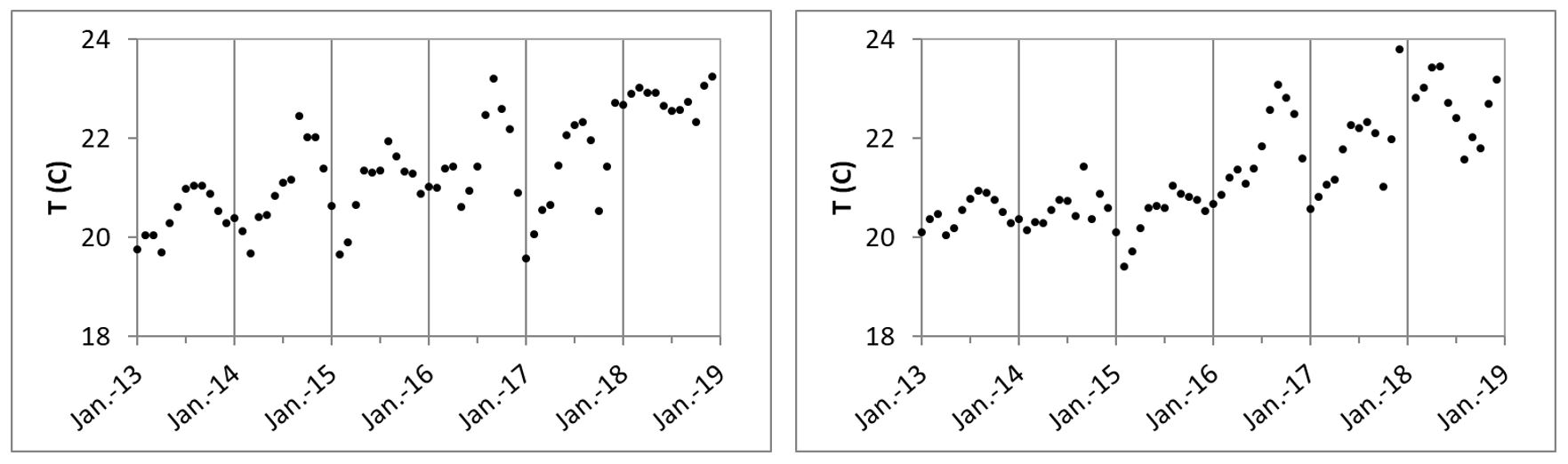}
\end{minipage}
\caption{Monthly average of the temperature in the hall B (left) and hall C (right) from January 2013 to December 2018. The sharp increase of the temperature due to ArDM experiment (see text) is very attenuated in the hall B and it is almost imperceptible in the hall C}
\label{fig:T-halls_ByC}
\end{figure*}

The $^{222}$Rn concentration in halls A, B and C are highly correlated, $r=0.94$ to 0.98, and similar behaviour can be found for the pressure and the relative humidity.
The temperature behaviour is different: there is correlation between the contiguous halls B and C, $r = 0.86$; the halls A and B are slightly correlated, $r = 0.53$; meanwhile the halls A and C are uncorrelated, $r = 0.32$.
This is a consequence of the liquefaction for several months \cite{Calvo_2017} of the argon needed for the ArDM experiment, that increased the temperature of the hall A (Figure~\ref{fig:datos-halla}) but not those of halls B and C (Figure~\ref{fig:T-halls_ByC}).
There is also correlation between the relative humidity and the $^{222}$Rn concentration, $r = 0.7$, for the nine pairs of the three halls. There exists a hint of a minor correlation between the relative humidity and the pressure, with $r$ close to 0.39. The other pairs are not significantly correlated.

\begin{figure}
\begin{minipage}{\columnwidth}
\centering
\includegraphics[width=1.00\textwidth]{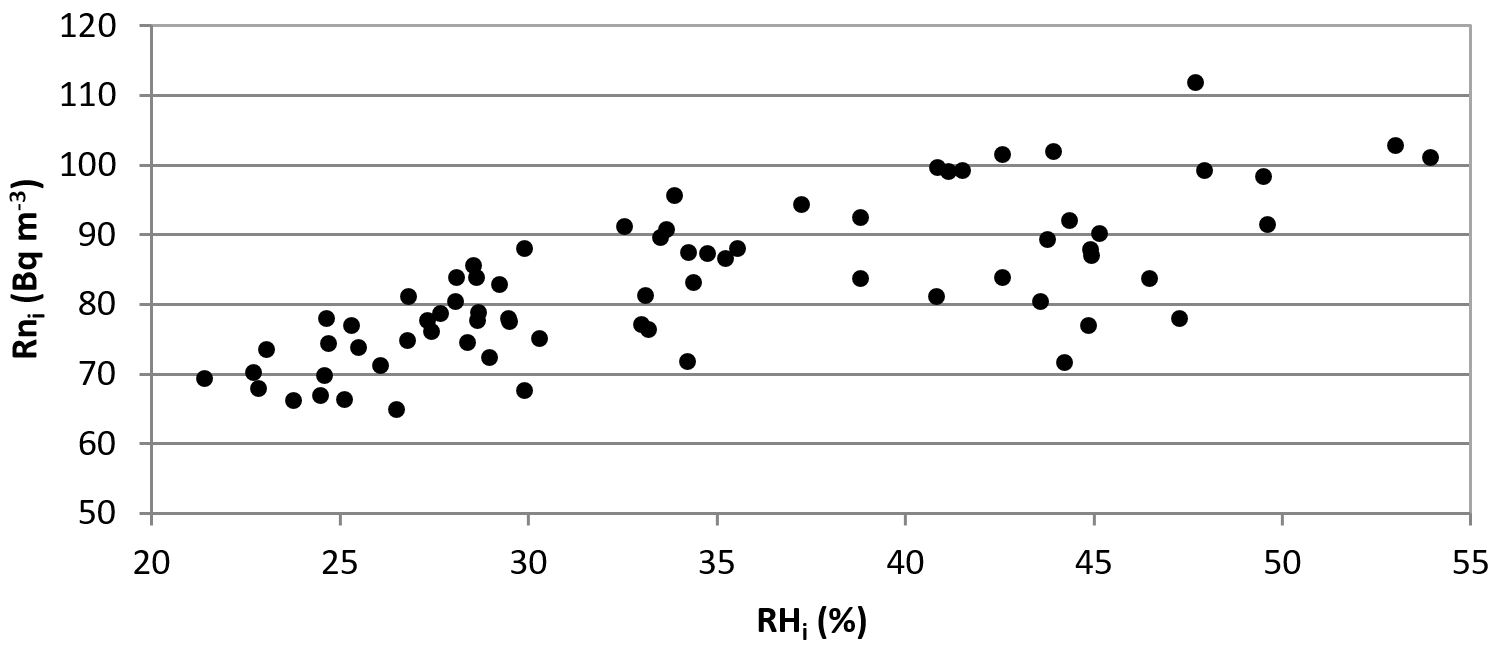}
\end{minipage}
\caption{Monthly averages of the $^{222}$Rn concentration versus the monthly averages of the relative humidity at hall A from 2013 to 2018. Corresponding plots for halls B and C are similar}
\label{fig:rn-hr-halla}
\end{figure}

Figure~\ref{fig:datos-halla} plots the monthly averages of $^{222}$Rn concentration, relative humidity, temperature and atmospheric pressure for the six years for the hall A. Halls B and C have very similar plots, except for the temperature (Figure~\ref{fig:T-halls_ByC}) because both halls were not very affected by the argon liquefaction of the ArDM experiment. The statistical uncertainty of the $^{222}$Rn monthly average is 0.24\% and those of the other three variables are unknown because of the lack of information in the data sheet of the radon monitor, though likely they are less than their calibration uncertainties.
The relative humidity has a clear seasonal periodicity and the $^{222}$Rn concentration points to a similar behaviour, more evident in 2013, 2015 and 2018. Both seasonal periodicities produce the here reported correlation between both parameters in the three halls (Table~\ref{tab:pearsonLSC} and Figure~\ref{fig:rn-hr-halla}). The seasonal periodicity is not observed either for the atmospheric pressure or the room temperature (Figure~\ref{fig:datos-halla}) because the former is determined by the external atmospheric pressure (non seasonal, see section~\ref{sec:aemet_data}) and the latter depends, mainly, on the internal air conditioning system and, sometimes, on considerable local alterations.

\subsection{Seasonal periodicity of the $^{222}$Rn concentration and of the relative humidity inside the laboratory}
\label{subsec:rn-rh-13-18}

The airborne radon depends on the radon concentration in soil gases and on its release to the atmosphere. A long-term study on the effects of climate on soil radon \cite{asher1992} found that wind speed, relative humidity, air and soil temperatures, and the difference between those temperatures cause both day-to-day and seasonal variations in soil radon, being seasonal variations of greater magnitude than day-to-day fluctuations. Although meteorological variables are not independent and it is difficult to estimate their relative impact, the authors of reference \cite{asher1992} were able to conclude that precipitation appears to be the most important meteorological variable affecting the content and the seasonal variation of $^{222}$Rn in soil gases. Differences between summer and winter $^{222}$Rn concentration have been reported in large dwellings samples \cite{arvela1995rpd,BOCHICCHIO2005rm}. The general trend is that indoor $^{222}$Rn is higher in winter than in summer (winter/summer ratio, W/S>1); we have observed in LSC the opposite result. However, the dispersion of the measurements in \cite{arvela1995rpd,BOCHICCHIO2005rm} is significant and there exist values of W/S<1. Other observations of W/S<1 have also been reported, see \cite{tanner1992} and references therein; on the other hand, the result of \cite{hans_lyon1986} shows that W/S>1 for indoor $^{222}$Rn, whereas W/S<1 for outdoor $^{222}$Rn. A quantitative model \cite{Arvela2016} provides a sinusoidal variation in the monthly average of the indoor $^{222}$Rn concentration with W/S>1. It is worth noting that this model has been applied with the typical parameters (outdoor--indoor temperature gradient, wind speed and $^{222}$Rn diffusion and convection rates) of Finnish climate and houses. But the model also shows that if the building materials are the dominant radon source, then the maximum $^{222}$Rn concentration occurs in summer.

The annual periodicity of the relative humidity and the concentration of $^{222}$Rn at LSC can be analyzed fitting the values of their monthly average to an equation of the type

\begin{equation}
	Y=A+B\cos \left(\frac{2\pi (t-t_{M})}{T}\right) \label{coseno}
\end{equation}

\noindent where $A$ and $B$ are the annual average and the modulation amplitude, respectively, $T$ is the period, $t_{M}$ is the time corresponding to the first maximum amplitude and $t$ is the measurement time after 1$^{\mathrm{st}}$ January 2013 at 00:00.

The parameters $A$, $B$, $T$ and $t_{M}$ are estimated by an unweighted least squares fit because the standard deviations of the measurements of the relative humidity are unknown and, although we know the statistical uncertainty of the radon concentration, we do not know its standard deviation due to its natural variability. Therefore, the square root of the minimum of the sum of the squared residuals per degree of freedom is an estimate of the standard deviation of a measurement, assuming that it is the same for each measurement \cite{bevington}. The results for the relative humidity are listed in Table~\ref{tab:rh_halls}.
The estimated standard deviations of the relative humidity monthly averages are 3.6\%, 4.2\% and 4.3\% for the halls A, B and C, respectively. The three periods are compatible with 1~year and the three maxima {$t_{M}$} are equal (within uncertainties), their mean value is ${215 \pm 3}$ d, around 3$^{\mathrm{rd}}$ August.

\begin{table}
\caption{Values of the fitted parameters of the relative humidity of the years 2013--2018 to Eq. (\ref{coseno}). The maxima are in the first week of August}
\label{tab:rh_halls}
\centering
\begin{tabular}{lcccc}
\hline
Room	&{A} (\%) & {B} (\%)	& {T} (d)	& {t$_{M}$} (d) \\
\hline
Hall A & ${34.6 \pm 0.4}$	& ${11.2 \pm 0.6}$ & ${364 \pm 2}$ & ${214 \pm 5}$ \\
Hall B & ${41.8 \pm 0.5}$ & ${11.7 \pm 0.7}$ & ${363 \pm 2}$ & ${217 \pm 6}$ \\
Hall C & ${40.2 \pm 0.5}$ & ${12.0 \pm 0.7}$ & ${364 \pm 2}$ & ${215 \pm 6}$ \\
\hline
\end{tabular}
\end{table}

The concentrations of $^{222}$Rn were also fitted to the Eq.~(\ref{coseno}). Table~\ref{tab:222rn_halls} shows the values of the fitted parameters. The unweighted least squares fits show that the estimated standard deviations of the measurements are 7.6, 7.6 and 8.0 Bq~m$^{-3}$ for the halls A, B and C, respectively. These values imply a mean standard deviation of 9\% for the $^{222}$Rn concentration monthly averages, a value much greater than their statistical uncertainty (0.24\%, see section~\ref{sec:Rn six years data}). Therefore, the observed $^{222}$Rn monthly deviation from annual modulation is not linked to the measurement precision, but to the natural variability of the $^{222}$Rn at the mountain where the fresh air is taken. 

\begin{table}
\caption{Values of the fitted parameters of the $^{222}$Rn concentration of the years 2013--2018 to Eq. (\ref{coseno}). The maxima are in mid-July}
\label{tab:222rn_halls}
\centering
\begin{tabular}{lcccc}
\hline
Room	&{A} (Bq~m$^{-3}$) & {B} (Bq~m$^{-3}$)	& {T} (d)	& {t$_{M}$} (d) \\
\hline
Hall A & ${82.8 \pm 0.9}$	& ${10.6 \pm 1.3}$ & ${366 \pm 4}$ & ${194 \pm 12}$ \\
Hall B & ${85.2 \pm 0.9}$ & ${10.0 \pm 1.3}$ & ${368 \pm 4}$ & ${193 \pm 13}$ \\
Hall C & ${84.8 \pm 0.9}$ & ${~~9.8 \pm 1.3}$ & ${367 \pm 5}$ & ${189 \pm 14}$ \\
\hline
\end{tabular}
\end{table}

\begin{figure}
\begin{minipage}{\columnwidth}
\centering
\includegraphics[width=1.00\textwidth]{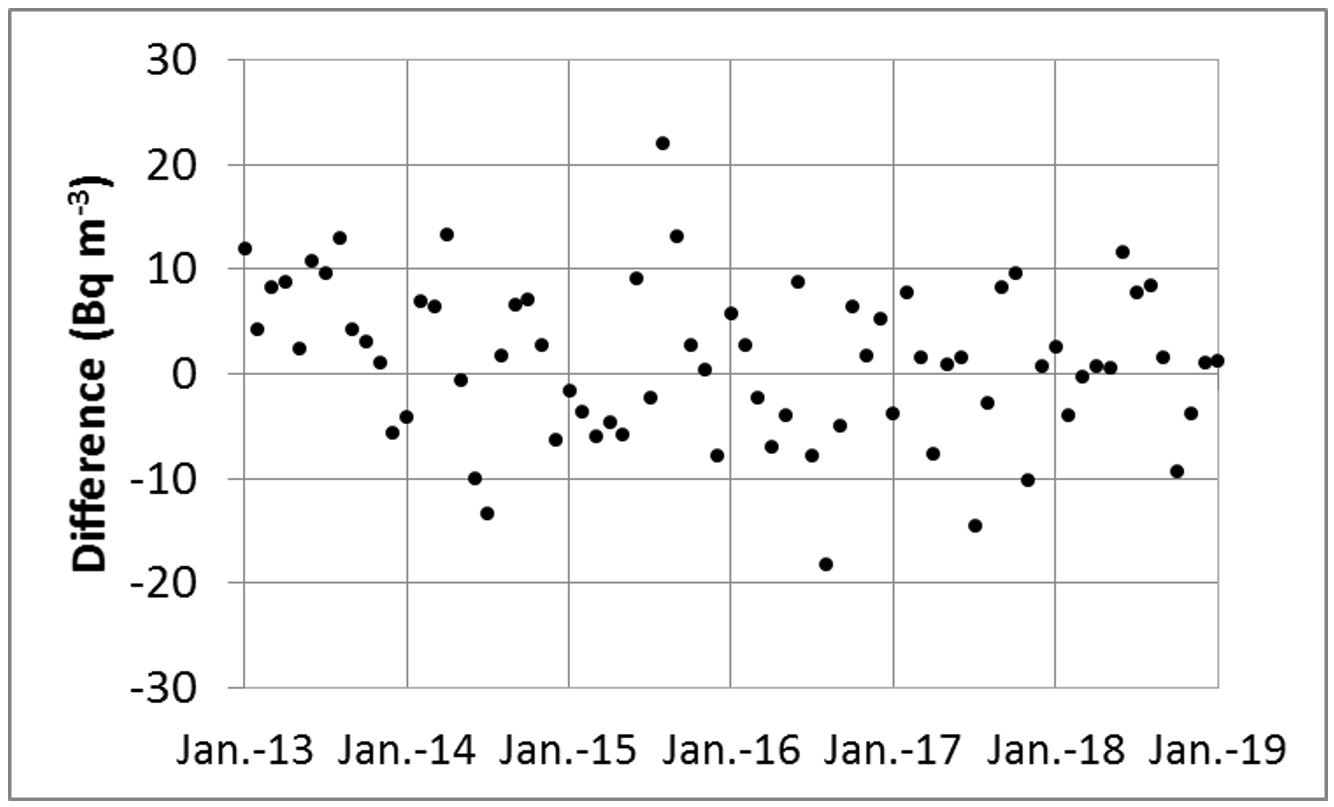}
\end{minipage}
\caption{Differences between the experimental and the fitted values of $^{222}$Rn concentration of hall A to Eq.~(\ref{coseno}). Corresponding plots for halls B and C are similar}
\label{fig:Rn_hallA_residuos}
\end{figure}

The differences between the observed and the fitted values do not show any particular trend (Figure~\ref{fig:Rn_hallA_residuos}). Most of the differences are less than 10~Bq~m$^{-3}$ and their maximum is 20~Bq~m$^{-3}$. Therefore, the sinusoidal function is a reasonable model. As in the case of the humidity, the three maxima, {$t_{M}$}, are compatible within their uncertainties, giving a mean value of ${192 \pm 8}$ d, around 11$^{\mathrm{th}}$ July. This means that the $^{222}$Rn maxima (minima) occur ${23 \pm 8}$ days before the humidity maxima (minima).

The origin of the winter-summer $^{222}$Rn level difference can be attributed to the different conditions of the airflow and weather patterns. There exist different features of the $^{222}$Rn concentration in different laboratories. For instance, the annual periodicity has not been observed in the Laboratori Nazionali del Gran Sasso (LNGS) \cite{BASSIGNANIrm1997}, which is connected to a road tunnel in Italy, under the Gran Sasso mountain. On the contrary, the periodicity has been observed in the Soudan Underground Laboratory (SUL) \cite{Goodmancrc1999}, located in an old iron mine in Minnesota, with the maximum in summer and the minimum in winter, as we report in LSC. However, as the relative humidity in SUL is constant, it is not correlated to the $^{222}$Rn concentration inside the laboratory \cite{PhysRevC.96.044609}.

It was suggested in \cite{Goodmancrc1999} that the radon periodicity in the SUL is due to the pressure gradients created by the relative temperature differences between the air in the laboratory and that on the surface. This qualitative model can also be applied to LSC: in winter, when the outside temperature is lower than the inside one, the ventilation rate is enhanced and the purge of the inside radon is more efficient than in summer, when the temperature gradient is reversed. Then, the $^{222}$Rn concentration has a positive correlation with the outside temperature and the winter-summer ratio, W/S, inside the laboratory is less than one. The observed positive correlation between the radon level and the relative humidity inside the LSC can be explained because in winter the rock of the tunnel contains less water than in summer, when the snow had melted mostly well before and the rock is plenty of water. In these conditions, the relative humidity inside the laboratory is greater in summer than in winter. Therefore, the high (low) humidity inside LSC is a consequence of the high (low) outside temperature.

\section{Meteorological data}
\label{sec:aemet_data}

A quantitative estimate of the positive correlation between the $^{222}$Rn concentration and the temperature outside the laboratory can be obtained from the data provided by the Spanish Meteorological Agency (Agencia Espa\~nola de Meteorología, AEMET).

The nearest AEMET observation point to the LSC is the Canfranc station, 3~km away from the underground laboratory. It records the temperature (T$_{\mathrm{o}}$) and relative humidity (RH$_{\mathrm{o}}$) every 10 minutes. The Arag\"u\'es del Puerto station, 13~km away, records the atmospheric pressure (P$_{\mathrm{o}}$) four times per day: midnight, 7:00, 12:00 and 18:00 hours. Table~\ref{tab:aemet_stations} lists the altitudes and coordinates of LSC and both AEMET stations.

\begin{table}
\caption{Altitude and coordinates of the LSC and of the AEMET stations used in this work. The stations data were provided by AEMET; those of LSC are our estimates on the application iberpix \cite{iberpix} of the Spanish National Geographic Institute (Instituto Geográfico Nacional)}
\label{tab:aemet_stations}
\centering
\begin{tabular}{ccccc}
\hline
Site & Altitude (m) & Latitude & Longitude \\
\hline
LSC & $1120$	& $42^\circ$ $46'$ $30''$ N & $0^\circ$ $31'$ $42''$ W \\
Canfranc & $1170$ & $42^\circ$ $44'$ $58''$ N & $0^\circ$ $30'$ $58''$ W \\
Arag\"u\'es del Puerto & $1040$ & $42^\circ$ $42'$ $32''$ N & $0^\circ$ $40'$ $23''$ W \\
\hline
\end{tabular}
\end{table}

We use data from 2015 to 2018 because the meteorological Canfranc data are available since the end of October, 2014. The correlation coefficients between the inside (four) and outside (three) studied variables from 2015 to 2018 for the three halls of the LSC are listed in Table~\ref{tab:pearsonLSC-AEMET}. 
In this case with 36 evaluated $r$--values, we use 99.86\% CL for the decision threshold because $1-(0.9986)^{36}=0.05$, see section \ref{sub:correla_inside}. The monthly average 4-year data gives $n=48$ in Eq. (\ref{student}), i.e., 46 degrees of freedom for the Student's $t$ and $\left|r_{\alpha=0.0014}\right|=0.46$.

\begin{table}
\caption{Correlation coefficients (2015-18) between the four inside variables, recorded underground by our radon monitors, for the halls A, B and C of LSC and the outside variables recorded by the AEMET stations of Canfranc (temperature and relative humidity) and Arag\"u\'es del Puerto (pressure)}
\label{tab:pearsonLSC-AEMET}
\centering
\begin{tabular}{lrrr}
\hline
\multicolumn{1}{c}{Room} & \multicolumn{3}{c}{Outside variables} \\
\hline
Hall A  & T$_{\mathrm{o}}$ & P$_{\mathrm{o}}$ & RH$_{\mathrm{o}}$ \\
\hline	
    Rn$_{\mathrm{i}}$ & 0.71 & 0.10 & $-$0.10 \\
    T$_{\mathrm{i}}$ & 0.14 & $-$0.01 & $-$0.20 \\
    P$_{\mathrm{i}}$ & 0.38 & 0.99 & $-$0.40 \\
    RH$_{\mathrm{i}}$ & 0.95 & 0.25 & $-$0.25 \\
\hline
Hall B  & T$_{\mathrm{o}}$ & P$_{\mathrm{o}}$ & RH$_{\mathrm{o}}$ \\
\hline
    Rn$_{\mathrm{i}}$ & 0.68 & 0.09 & $-$0.08 \\
    T$_{\mathrm{i}}$ & 0.28 & $-$0.14 & 0.32 \\
    P$_{\mathrm{i}}$ & 0.32 & 0.97 & $-$0.36 \\
    RH$_{\mathrm{i}}$ & 0.94 & 0.29 & $-$0.34 \\
\hline
Hall C  & T$_{\mathrm{o}}$ & P$_{\mathrm{o}}$ & RH$_{\mathrm{o}}$ \\
\hline
    Rn$_{\mathrm{i}}$ & 0.64 & 0.09 & $-$0.16 \\
    T$_{\mathrm{i}}$ & 0.10 & $-$0.15 & 0.31 \\
    P$_{\mathrm{i}}$ & 0.39 & 0.99 & $-$0.40 \\
    RH$_{\mathrm{i}}$ & 0.94 & 0.28 & $-$0.36 \\
\hline
\end{tabular}
\end{table}

\begin{figure}
\begin{minipage}{\columnwidth}
\centering
\includegraphics[width=1.00\textwidth]{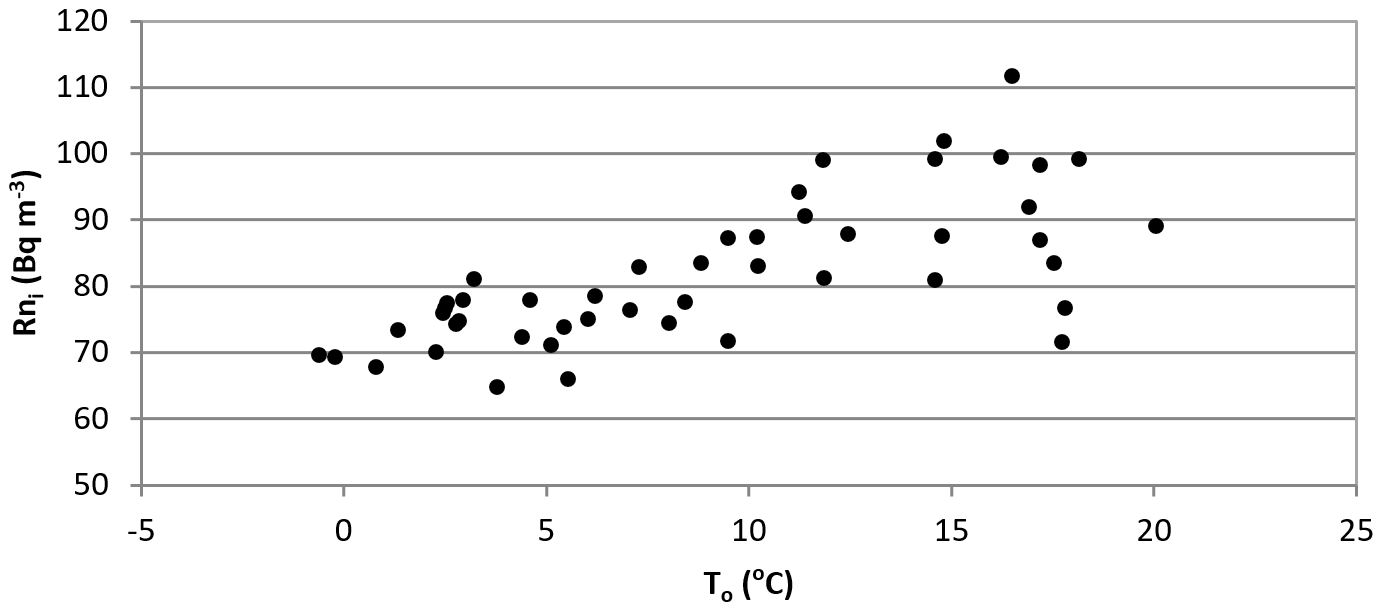}
\end{minipage}
\caption{Monthly averages of the $^{222}$Rn concentration at hall A versus the monthly averages of the outside temperature from 2015 to 2018. Corresponding plots for halls B and C are similar}
\label{fig:rn-hallA-Tout}
\end{figure}

\begin{figure*}
\begin{minipage}{\columnwidth}
\centering
\includegraphics[width=2.00\textwidth]{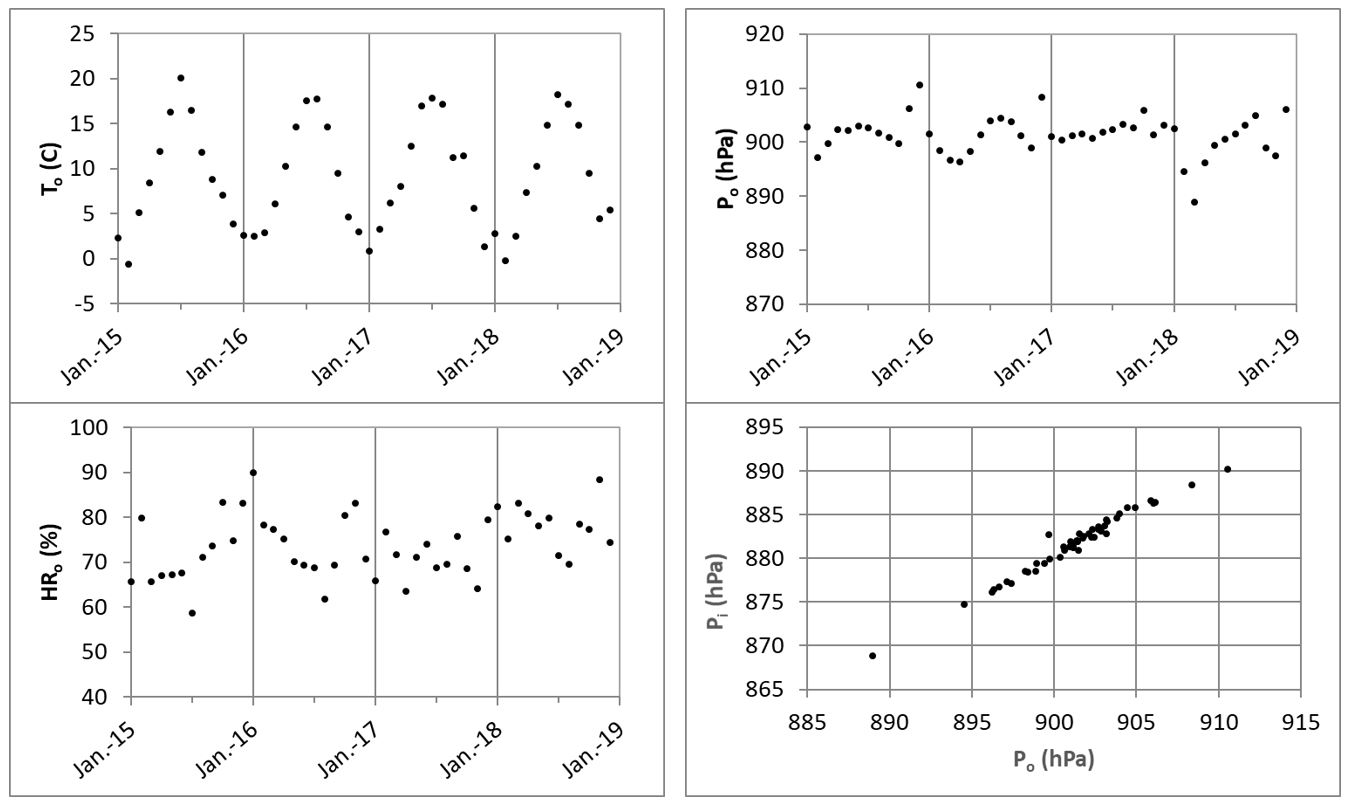}
\end{minipage}
\caption{Monthly averages of AEMET meteorological data: temperature (upper left), pressure (upper right) and relative humidity (bottom left). The bottom right panel shows the high correlation between the atmospheric pressures in hall A (P$_{\mathrm{i}}$) and Arag\"u\'es del Puerto (P$_{\mathrm{o}}$); halls B and C have similar correlation with P$_{\mathrm{o}}$}
\label{fig:outside}
\end{figure*}

Independently of the hall, we observe three main features regarding the correlations inside-outside.
First, the atmospheric pressures inside and outside are the variables with the highest correlation, $r\approx1$, because two close points (13 km away and 80 m altitude difference) have the same pressure gradient.
Second, the inside relative humidity and the outside temperature are highly correlated, $r = 0.94$, most likely linked to the thaw season because the outside temperature follows a sinusoidal curve of one-year period (Figure~\ref{fig:outside}) driving the inside relative humidity (Figure~\ref{fig:datos-halla}). Note also that 
the outside relative humidity (Figure~\ref{fig:outside}) is not a sinusoidal curve similar to the inside relative humidity.
Third, the inside $^{222}$Rn concentration is correlated with the outside temperature (Figure~\ref{fig:rn-hallA-Tout}), $r = 0.64$ to $r = 0.71$, a similar amount to the correlation between the inside $^{222}$Rn concentration and the inside relative humidity, $r = 0.67$ to $r = 0.73$ (Table~\ref{tab:pearsonLSC}). Latter correlation occurs in the same hall and also for different halls.

The Table~\ref{tab:LSC_AEMET} collects the results of fitting T$_{\mathrm{o}}$, RH$_{\mathrm{i}}$, and Rn$_{\mathrm{i}}$ to Eq.~(\ref{coseno}) within the same interval of available data for T$_{\mathrm{o}}$ (2015 to 2018). T$_{\mathrm{o}}$ has a period compatible with 1 year and the maximum is ${204 \pm 3}$ d, around 23$^{\mathrm{rd}}$ July. The period of the internal relative humidity, RH$_{\mathrm{i}}$, is also compatible with one year and the maxima {$t_{M}$} in every hall are equal within uncertainties, their mean value is ${213 \pm 3}$ d, around 1$^{\mathrm{st}}$ August, ${9 \pm 5}$ days after the maximum of T$_{\mathrm{o}}$. Though this delay has a low statistical significance, it could be a rough estimate of the time interval between the maximum thaw on the Tobazo Mountain and the maximum rock humidity at 800~m rock depth. The mean values and modulation amplitudes of the internal relative humidity are similar to those of the 2013-2018 series (Table~\ref{tab:rh_halls}).
The $^{222}$Rn results of the 2015-2018 series show three periods compatible with one year and three maxima ${10\pm14}$ days after the maximum of T$_{\mathrm{o}}$.
Compared to the 2013-2018 series (Table~\ref{tab:222rn_halls}), the subset 2015-2018 exhibits differences in the estimated parameters (the most remarkable are $\sim10$ days in the period values and $\sim3$ weeks in the {t$_{M}$} values). In this subset the $^{222}$Rn maxima (minima) occur at the same time the humidity maxima (minima), whereas in the total set 2013-2018 they occur ${23 \pm 8}$ days before (see section \ref{subsec:rn-rh-13-18}). This behaviour appears because the seasonal periodicity of the $^{222}$Rn concentration has more fluctuations than the one of the relative humidity, see the fits of the Figure~\ref{fig:datos-halla}, because it is more sensitive to variations in the environmental conditions; for instance, the outside rainfall changes the radon content of the fresh air cleaning the laboratory, but it does not change the relative humidity inside the laboratory, 800 meters down in the rock.

\begin{table}
\caption{Values of the fitted parameters (2015-18) according Eq.~(\ref{coseno}) of the variables showing seasonal periodicity: outside temperature (maximum on 23$^{\mathrm{rd}}$ July), inside relative humidity and inside radon concentration (maxima on beginning August)}
\label{tab:LSC_AEMET}
\centering
\begin{tabular}{lcccc}
\hline
\multicolumn{5}{c}{Outside temperature}  \\
          &{A} ($^\circ$C) & {B} ($^\circ$C)	& {T} (d)	& {t$_{M}$} (d) \\
\hline
Canfranc & ${9.1 \pm 0.2}$	& ${8.1 \pm 0.3}$ & ${366 \pm 2}$ & ${204 \pm 3}$ \\
\hline
\multicolumn{5}{c}{Inside relative humidity}  \\
       &{A} (\%) & {B} (\%)	& {T} (d)	& {t$_{M}$} (d) \\
\hline
Hall A & ${33.9 \pm 0.4}$	& ${11.1 \pm 0.5}$ & ${361 \pm 2}$ & ${218 \pm 4}$ \\
Hall B & ${40.5 \pm 0.5}$ & ${11.9 \pm 0.7}$ & ${365 \pm 3}$ & ${210 \pm 6}$ \\
Hall C & ${39.3 \pm 0.5}$ & ${12.2 \pm 0.7}$ & ${367 \pm 3}$ & ${208 \pm 6}$ \\
\hline
\multicolumn{5}{c}{Inside $^{222}$Rn concentration}  \\
       &{A} (Bq~m$^{-3}$) & {B} (Bq~m$^{-3}$)	& {T} (d)	& {t$_{M}$} (d) \\
\hline
Hall A & ${82.0 \pm 1.1}$	& ${11.2 \pm 1.5}$ & ${357 \pm 7}$ & ${214 \pm 13}$ \\
Hall B & ${84.9 \pm 1.1}$ & ${10.4 \pm 1.5}$ & ${358 \pm 7}$ & ${215 \pm 14}$ \\
Hall C & ${84.4 \pm 1.1}$ & ${10.3 \pm 1.5}$ & ${358 \pm 8}$ & ${214 \pm 14}$ \\
\hline
\end{tabular}
\end{table}

\section{$^{222}$Rn control in the ANAIS--112 experiment}
\label{sec:Rn in anais}

The $^{222}$Rn annual periodicity in the Hall B of LSC was also reported by the ANAIS collaboration \cite{maolivan2016}, an experiment looking for the dark matter annual modulation with NaI(Tl) scintillators at LSC \cite{Amare2021prd,Amare2019prl,Amare:2018sxx,Amare2019epjc} to test the DAMA-LIBRA result \cite{dama2020ijmpa} using the same target and experimental approach. To keep low the concentration of $^{222}$Rn, the air inside the ANAIS--112 shielding has been kept continuously flushed with radon-free nitrogen gas. During most of the\linebreak[4] ANAIS--112 operation this radon-free nitrogen gas has been produced by evaporating liquid nitrogen, but also pure nitrogen from compressed gas bottles has been used.
Although there was no direct measurement of the $^{222}$Rn content of this nitrogen, from the absence of lines coming from $^{214}$Bi and $^{214}$Pb in the current NaI(Tl) detectors \cite{Amare2016epjc} as well as in the ANAIS--0 prototype \cite{CEBRIAN2012app}, the ANAIS collaboration estimated an upper limit of 0.6 Bq~m$^{-3}$ of $^{222}$Rn in the air filling the shielding.

\begin{figure}
\begin{minipage}{\columnwidth}
\centering
\includegraphics[width=1.00\textwidth]{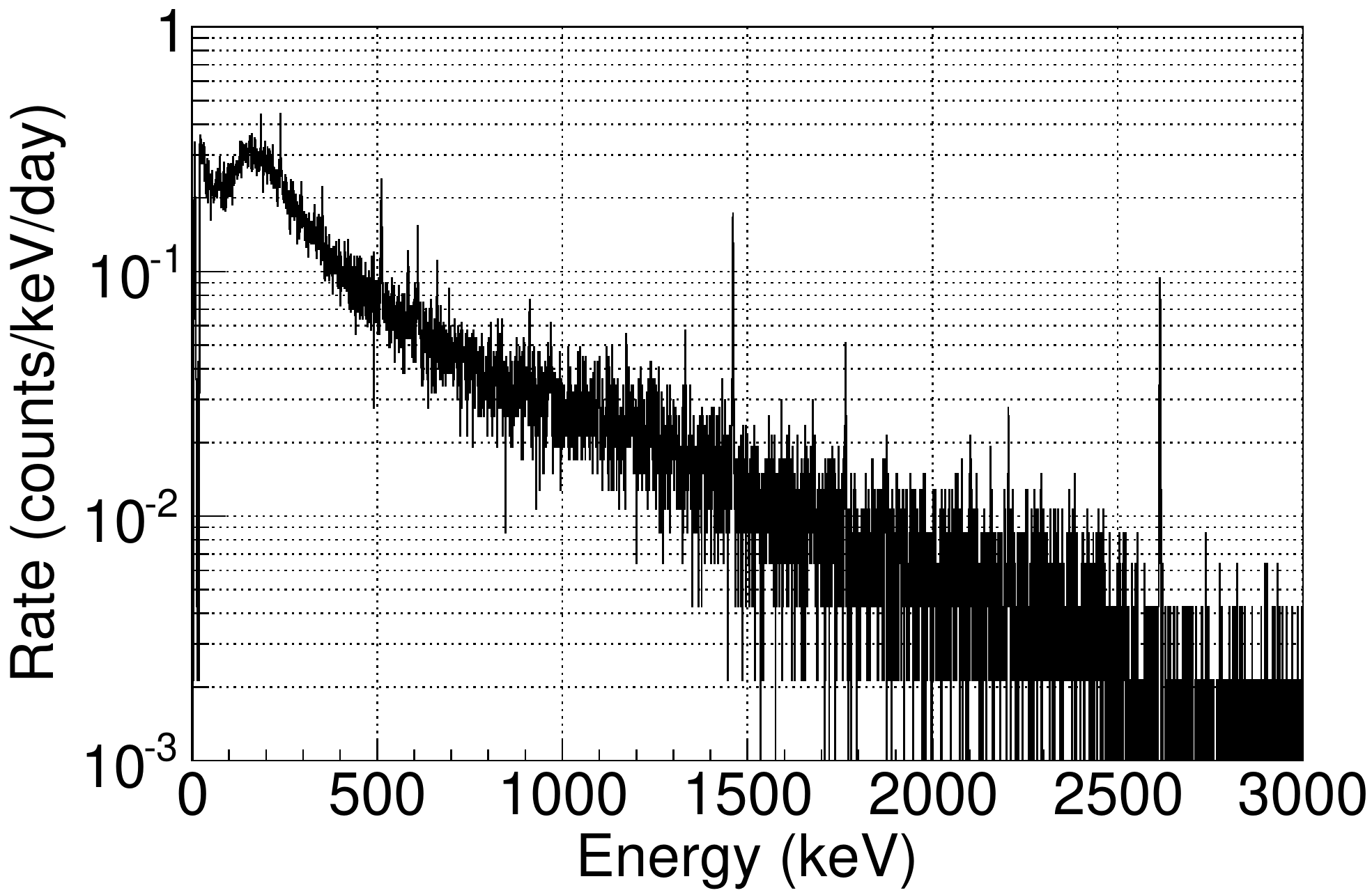}
\end{minipage}
\caption{Total HPGe spectrum for 2017-2018 (670 days live time)}
\label{fig:spc670d}
\end{figure}

This upper limit can be improved through HPGe gamma spectrometry. We purged for two years (2017-2018) the inner volume ($20\times20\times30$~cm$^{3}$) of the shielding \cite{Coarasa2016jpcs} of a HPGe detector ($\sim$1~kg and 41\% relative efficiency with respect to a 3~in$\times$3~in NaI(Tl) detector) with the same gas purging the ANAIS--112 shielding. The gain shifts were controlled weekly with a $^{22}$Na source. The total live time of the measurement was 670 days; the lost acquisition days were due to maintenance works in LSC, power outages and, in a minor amount, to the periodic calibrations and the refilling of the HPGe dewar with liquid nitrogen. The total spectrum is plotted in Figure~\ref{fig:spc670d}, where the annihilation peak at 511~keV and the peaks at 1460.8~keV ($^{40}$K) and at 2614.6~keV ($^{208}$Tl) are clearly distinguishable.

\subsection{HPGe photopeak efficiency of an extended $^{222}$Rn source}

We have a precise knowledge of the efficiency of the HPGe used in this work because it is one of our ultra low background HPGe detectors located in the LSC, that is routinely used for the screening of material radiopurity. For this reason, its efficiency has been measured for samples facing the HPGe window and also for Marinelli beakers. The nitrogen source is quite different from the habitual samples because it fills all the internal cavity of the shielding and, therefore, the efficiency measurement cannot be completed for the total volume because the access to the rear of the HPGe is not possible once the shielding has been built.

\begin{figure}
\begin{minipage}{\columnwidth}
\centering
\includegraphics[width=1.00\textwidth]{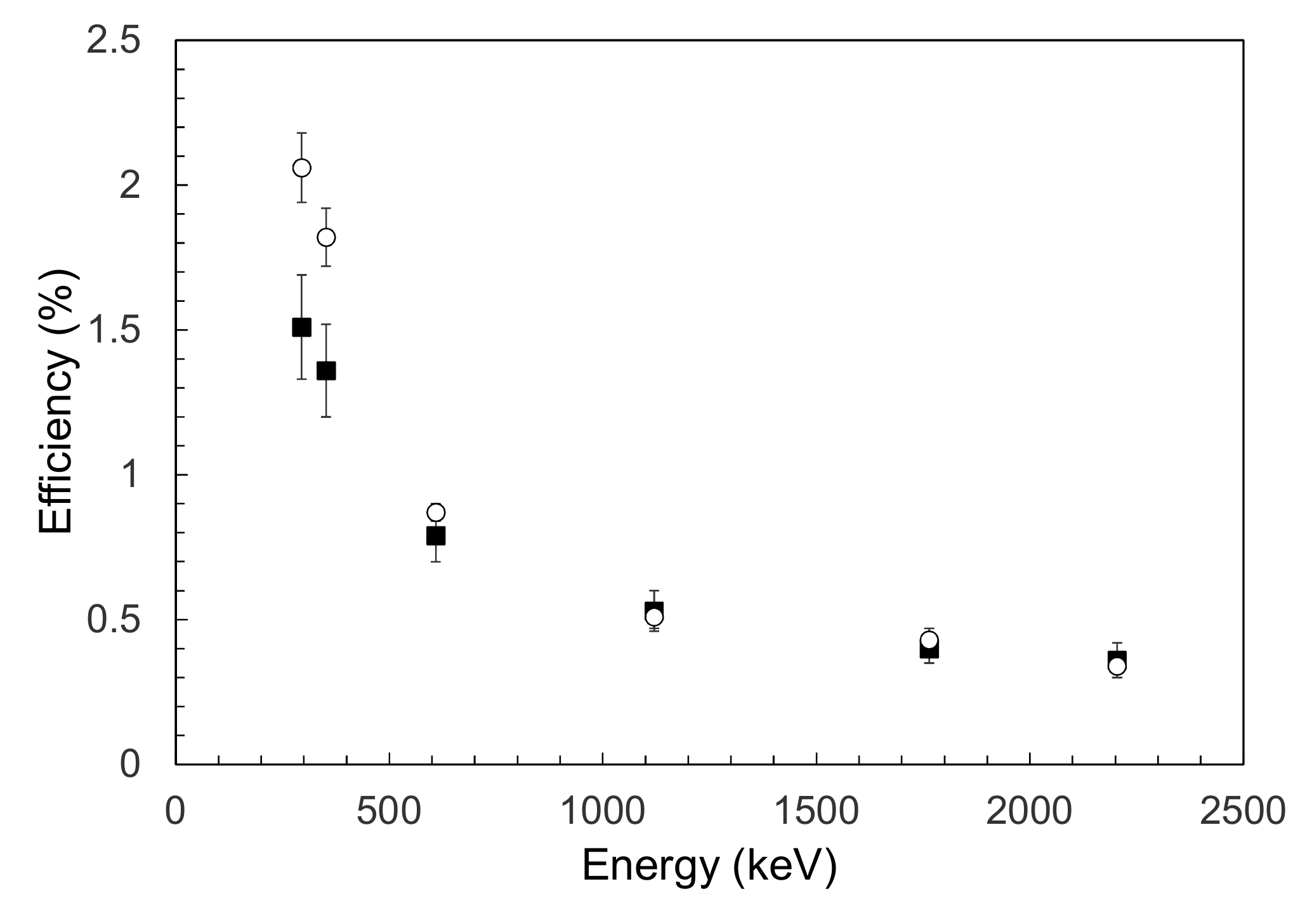}
\end{minipage}
\caption{Efficiency of our HPGe detector to a $^{222}$Rn source filling the inner volume of its shielding. The squares were estimated profiting a $^{222}$Rn efficiency relative measurement and the circles using Geant4 with an approximate geometry. In both cases the estimated values were normalized to the measured efficiency at 1460.8 keV}
\label{fig:eficiencias}
\end{figure}

The efficiency to the $^{222}$Rn source was estimated in two steps. First, we measured the efficiency at 1460.8~keV,\linebreak[4] $\epsilon(1460.8$~keV), using a homemade source of 260~g of KCl ($^{40}$K isotopic abundance of 0.0117\% \cite{Browne1986}) and chemical purity >95\%, located at several positions in the accessible volume around the HPGe, and a smaller one of 20.6~g to measure the HPGe efficiency at several distances from the HPGe window. The self-attenuation of the 1460.8~keV gamma ray is 10\% for the 260~g source and 2\% for the 20.6~g source. 
Second, we used the airborne $^{222}$Rn filling the inner volume of the shielding due to an accidental cut of the purging system. Because of the activity of $^{222}$Rn was unknown, we were able to obtain only the relative efficiency, $\epsilon_{R}(E)$, of the HPGe to a source identical to the one we intended to measure.

A linear interpolation, in a double logarithmic scale, between $\epsilon_{R}(1120.3$~keV) and $\epsilon_{R}(1764.5$~keV) was used to calculate the relative efficiency $\epsilon_{R}(1460.8$~keV). We take into account in this interpolation the true coincidence summing of 1120.3 and 609.3~keV through a correction factor\linebreak[4] $c_{TS}\sim1.1$ for $\epsilon_{R}(1120.3$~keV), that we estimated by Monte Carlo for the extended $^{222}$Rn source. The matching factor $f_{m}=\epsilon(1460.8$~keV)$/\epsilon_{R}(1460.8$~keV) gives the efficiencies for the $^{222}$Rn gamma lines (Figure~\ref{fig:eficiencias}).

The matching factor $f_{m}$ has a main uncertainty linked to the lack of measurements for sources located at the rear of the HPGe, where we know neither the materials nor the sizes of the detector components (cooling rod, electronics, supports, etc.). Initially we assumed that the HPGe efficiencies to rear and frontal sources were identical, but if the detector components attenuated all the photons coming from the rear volume, then $f_{m}$ would decrease a 20\%. Hence, we have assumed that the rear contribution is one half of the frontal one, we have diminished $f_{m}$ by a 10\% and we have added a 10\% systematic uncertainty to all the values of the Figure~\ref{fig:eficiencias}.

As a crosschecking, we did a Monte Carlo simulation with Geant4 \cite{geant4nim2003} according to the known dimensions and materials of the germanium crystal and its end cup.
The results of the simulation were multiplied by a factor 0.94 to match the experimental and calculated efficiencies at\linebreak[4] 1460.8~keV for the 20.6~g source located in frontal positions. Subsequently, the same factor was applied to the calculated efficiencies of the more intense gamma lines of the $^{222}$Rn daughters (Figure~\ref{fig:eficiencias}) coming from the inner volume of the HPGe shielding.
This estimate agrees with the former around the normalization energy (1460.8~keV), as expected, and shows some differences at lower energies because a more precise knowledge of the detector geometry is needed to match by Monte Carlo the efficiencies at 295.2 and 351.9~keV.

\subsection{Upper limit to $^{222}$Rn concentration}

The observed peaks of the $^{222}$Rn progeny can also be produced by $^{226}$Ra and $^{238}$U contamination in the shielding or in the HPGe itself, but our current data cannot separate the three contributions. Then, we only can estimate an upper limit to the residual $^{222}$Rn in air. The best upper limit, 0.06 Bq~m$^{-3}$ (95\% C.L.), is obtained with the 351.9~keV gamma line. This value is one order of magnitude better than our previous estimate based on the NaI(Tl) spectrometry data and it is not affected by the systematic uncertainty of $f_{m}$ because the dominant one is statistical.
According to the ANAIS--112 background model \cite{Amare2016epjc}, this upper limit means that the contribution of the $^{222}$Rn to the total rate is\linebreak[4] $<4\times10^{-4}$ counts keV$^{-1}$kg$^{-1}$d$^{-1}$ in the region of interest, [1--6]~keV, amounting to less than 0.01\% of the background \cite{Amare2021prd}.

The possible contribution of some modulation of some tiny $^{222}$Rn activity to the expected signal in ANAIS--112 is negligible because if we would take the above upper limit of 0.06 Bq~m$^{-3}$ as a true $^{222}$Rn concentration with a modulation amplitude $\sim10\%$, similar to the one observed in the Hall~B (Table \ref{tab:222rn_halls}), its contribution to the ANAIS--112 modulation signal would be $4\times10^{-5}$ counts~keV$^{-1}$kg$^{-1}$d$^{-1}$ in the region [1--6]~keV, much below the modulation observed by the DAMA/LIBRA collaboration \cite{dama2020ijmpa} and the sensitivity of ANAIS--112 experiment \cite{Coarasa2019epjc}.

\section{Upper limit to $^{222}$Rn concentration of the radon-free air provided by the LSC}

The laboratory has installed a radon abatement system (RAS), built by the company Apparatuses Technologies Construction (ATEKO a. s.) from Czech Republic, that can provide 220~m$^{3}$h$^{-1}$ with $\sim$1~mBq~m$^{-3}$ of $^{222}$Rn for the experiments hosted by the LSC   \cite{perezperez2021radon}. Though this $^{222}$Rn concentration is very low to be detected by our HPGe detector, we have tested the $^{222}$Rn content from the RAS as we did for the nitrogen purging ANAIS--112 (see section \ref{sec:Rn in anais}). We have collected data for 25 months from January, 2019 to June, 2021; adding up to 728 days live time. There are two gaps in the data taking, from September to December, 2019 and April, 2021, that were devoted to other measurements. The remaining lost days are due to the weekly maintenance of the HPGe detector.

The Table~\ref{tab:compared_data} shows the total rate and the rates of the more intense lines of $^{222}$Rn daughters for both measurements. We have not observed significant differences between this measurement and the one done with evaporated nitrogen (section~\ref{sec:Rn in anais}). The total spectrum is equal, within statistical uncertainties, to the one obtained purging with nitrogen (Figure \ref{fig:spc670d}).

\begin{table}
\caption{Total rate from 50~keV to 5.6~MeV and rates of the more intense lines of $^{222}$Rn daughters for the two radon purging systems}
\label{tab:compared_data}
\centering
\begin{tabular}{ccc}
\hline
keV & Evaporated N$_{2}$ (d$^{-1}$) & RAS (d$^{-1}$) \\
\hline
    [50, 5600] & ${128.93 \pm 0.44}$ & ${129.63 \pm 0.42}$ \\
		~~295.2 & ${0.119 \pm 0.035}$ & ${0.116 \pm 0.032}$ \\
		~~351.9 & ${0.189 \pm 0.036}$ & ${0.216 \pm 0.033}$ \\
		~~609.3 & ${0.181 \pm 0.032}$ & ${0.223 \pm 0.029}$ \\
		1120.3 & ${0.051 \pm 0.023}$ & ${0.062 \pm 0.017}$ \\
		1764.5 & ${0.117 \pm 0.017}$ & ${0.079 \pm 0.015}$ \\
		2204.2 & ${0.052 \pm 0.012}$ & ${0.064 \pm 0.012}$ \\
\hline
\end{tabular}
\end{table}

The 351.9~keV line gives the best upper limit to the $^{222}$Rn around our HPGe detector, 0.06 Bq~m$^{-3}$ (95\% C.L.), the same value obtained when purging with evaporated nitrogen.

\section{Conclusions}

The measurements performed at the LSC over six years\linebreak[4] (2013-2018) have shown that the monthly average $^{222}$Rn concentration in air is lower than the maximum allowed for working places, 600 Bq~m$^{-3}$, according to the Spanish legislation \cite{csn11-04} and the international recommendations \cite{icrp103}. It is essentially similar in the three experimentation halls at LAB2400 (see Table \ref{tab:222rn_annos_mean}), indicating that the heating and the forced ventilation of the air reaching the LSC halls work properly.

There is a correlation between the $^{222}$Rn concentration and the relative humidity in the three halls.
An annual and sinusoidal periodicity of the values of the relative humidity and the $^{222}$Rn concentration was observed in the three halls.
The data from 2013 to 2018 show that the maximal humidity inside LSC occurs at beginning August and the maximal $^{222}$Rn concentration occurs 3 weeks before, at mid-July. The partial data from 2015 to 2018 show that the time of maximal humidity remains stable whereas the time of maximal $^{222}$Rn concentration is less stable, in fact, for this series it occurs at beginning August.
The relative amplitude of the $^{222}$Rn concentration with respect to the annual average is about 10\% for the three halls.
According to the AEMET data used in this work, the $^{222}$Rn concentration is correlated with the temperature of the fresh air intake on the mountain covering the LSC. The correlation with the relative humidity inside LSC is indirect, because of the likely correlation between the external temperature and the internal humidity via the snow melting.

The low $^{222}$Rn content (if any) of the nitrogen purging the ANAIS--112 shielding does not affect to the achieved low background and it guarantees the sensitivity of the experiment \cite{Coarasa2019epjc}.

We have also tested the radon abatement system of the laboratory in the same way we tested the ANAIS--112 purging system with evaporated nitrogen. We have obtained the same upper limit to the $^{222}$Rn concentration in both cases, 0.06 Bq~m$^{-3}$ (95\% C.L.), concluding that both systems are equivalent at this level of $^{222}$Rn sensitivity. 

\begin{acknowledgements}
This research was funded by\linebreak[4] MCIN/AEI/10.13039/501100011033 under grant PID2019-104374GB-I00; by MINECO-FEDER under grants FPA2017-83133-P, and\linebreak[4] FPA2014-55986-P; by MICINN-FEDER under grants FPA2011-23749; by CONSOLIDER-Ingenio 2010 Programme under grants MultiDark CSD2009-00064 and CPAN CSD2007-00042; by the University of Zaragoza under grant UZ2017-CIE-09; by the Spanish Meteorological Agency (AEMET), the Gobierno de Arag\'on (Group in Nuclear and Astroparticle Physics, ARAID Foundation and I. Coarasa predoctoral grant), the European Social Fund and by the LSC consortium. The authors would like to acknowledge the use of Servicio General de Apoyo a la Investigaci\'on-SAI, Universidad de Zaragoza. The authors also thank Carlos Pe\~na--Garay, Director of the LSC, Aldo Ianni and Alessandro Bettini (former LSC Directors) and Jos\'e \'Angel Villar (former LSC Associate Director) for their support and encouragement.
\end{acknowledgements}

\bibliographystyle{spphys}
\interlinepenalty=10000
\bibliography{radon_LSC}

\end{document}